\documentclass[12pt]{article}
\usepackage[margin=1.9cm]{geometry}
\usepackage{graphicx}
\usepackage{cite}

\newcommand{\mysection}{\setcounter{equation}{0}\section}

\def\beq{\begin{equation}}
\def\eeq{\end{equation}}
\def\beqa{\begin{eqnarray}}
\def\eeqa{\end{eqnarray}}
 
\newenvironment{Presented}{\begin{quotation} \begin{center} 
             PRESENTED AT\end{center}\bigskip 
             \begin{center}\begin{large}}{\end{large}\end{center} \end{quotation}}

\begin{document}

\begin{center}
{\Large \bf Higher-order corrections for $tqZ$ production}
\end{center}
\vspace{2mm}
\begin{center}
{\large Nikolaos Kidonakis$^{a}$ and Nodoka Yamanaka$^{b,c}$}\\
\vspace{2mm}
${}^a${\it Department of Physics, Kennesaw State University, \\ Kennesaw, GA 30144, USA} \\
\vspace{1mm}
${}^b${\it Kobayashi-Maskawa Institute for the Origin of Particles and the Universe, Nagoya University, \\ Furocho, Chikusa, Aichi 464-8602, Japan} \\
\vspace{1mm}
${}^c${\it Nishina Center for Accelerator-Based Science, RIKEN, \\ Wako 351-0198, Japan}
\end{center}

\begin{abstract}
We present theoretical results for the associated production of a single top quark and a $Z$ boson ($tqZ$ production) at LHC energies. We calculate higher-order corrections from soft-gluon emission for this process. We compute the approximate NNLO (aNNLO) cross section at LHC energies, including uncertainties from scale dependence and from parton distributions. We also calculate the top-quark rapidity distribution. The aNNLO corrections are significant and enhance the NLO cross section, and their inclusion provides a more precise theoretical prediction.
\end{abstract}
\vfill
\begin{Presented}
DIS2023: XXX International Workshop on Deep-Inelastic Scattering and
Related Subjects, \\
Michigan State University, USA, 27-31 March 2023 \\
     \includegraphics[width=9cm]{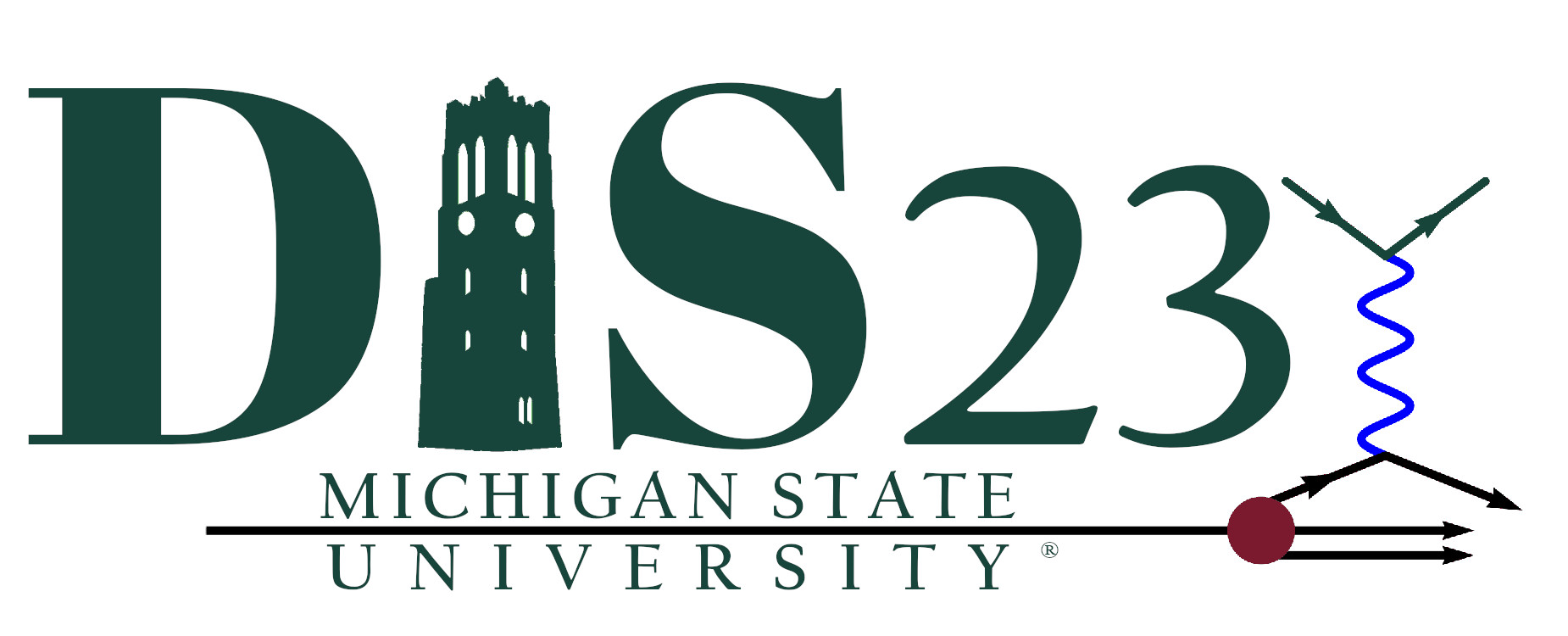}
\end{Presented}
\vfill

\newpage

\mysection{Introduction}

Single top-quark production in association with a $Z$ boson, i.e. $tqZ$ production with $q$ a light quark or antiquark, has been studied at the LHC for some time \cite{CMS1,ATLAS1,CMS2,CMS3,ATLAS2,CMS4}, leading to the observation of this process in $pp$ collisions at 13 TeV energy \cite{CMS3,ATLAS2}. Therefore, it is important to have improved theoretical calculations for this process. 

The cross section for $tqZ$ production allows the investigation of $t$-$Z$ and $W$-$W$-$Z$ couplings, and it is sensitive to anomalous top-quark couplings, for example, $t$-$q$-$Z$ \cite{NKAB,LZLGZ,NKtZ,MGNK,YLSM} and $t$-$q$-$g$ \cite{NKEM}. The $tqZ$ processes also probe the anomalous weak moments of the top quark which may be generated by new particles appearing through loop diagrams in a similar way as the anomalous electromagnetic moments \cite{CDVM,Yamanaka2017}.
Thus, $tqZ$ production provides good probes of new physics beyond the Standard Model.

The QCD corrections are significant at NLO \cite{CER}, and they are needed for good theoretical predictions since they provide an enhancement of around 16\% to the LO cross section at current LHC energies, while electroweak corrections \cite{PTV,DPS} are small, around 1\%. Further improvement in theoretical accuracy can be achieved by the inclusion of soft-gluon corrections, which are very important for top-quark processes \cite{NKsingletop,NK3loop,NKNYtW,NKttbar,FK2020,FK2021,NKNYtqgamma,NKNYtqZ,NKAT,NKloops} due to large contributions from soft-gluon emission near partonic threshold. This is well known for many $2 \to 2$ top-quark processes, including single-top production \cite{NKsingletop,NK3loop,NKNYtW} and top-antitop pair production \cite{NKttbar}, as well as some $2 \to 3$ processes \cite{FK2020}, in particular $tqH$ production \cite{FK2021}, $tq\gamma$ production \cite{NKNYtqgamma}, and $t{\bar t}\gamma$ production \cite{NKAT}. Here, we derive approximate NNLO (aNNLO) predictions for $tqZ$ production \cite{NKNYtqZ} by adding soft-gluon corrections at NNLO to the complete NLO results. 

In Sec. 2 we give a brief overview of soft-gluon resummation in $tqZ$ production. In Sec. 3 we present total cross sections for $tqZ$ and ${\bar t}qZ$ processes through aNNLO. In Sec. 4 we present top-quark rapidity distributions though aNNLO. We give a brief summary in Sec. 5. 

\mysection{Soft-gluon corrections}

We study hadronic processes $pp \to tqZ$. The partonic processes at LO are $a(p_a)\, + \, b\, (p_b) \rightarrow t(p_t) +q(p_q) + Z(p_Z)$. If an additional gluon is emitted with momentum $p_g$ in the final state, then we define the variable $s_4=(p_q+p_Z+p_g)^2-(p_q+p_Z)^2$. At partonic threshold, $s_4 \rightarrow 0$. Soft-gluon corrections appear at each order in the perturbative series in the form $[\ln^k(s_4/m_t^2)/s_4]_+$ with $k \le 2n-1$ for the order $\alpha_s^n$ corrections, where $m_t$ is the top-quark mass. The factorization of these soft-gluon corrections into a soft function in the cross section results in their resummation. The soft anomalous dimension $\Gamma_{\! S \, a b \rightarrow tqZ}$ controls the evolution of the soft function.

Next, we provide a short derivation of the resummed cross section. We first write the differential hadronic cross section, $d\sigma$, as a convolution of parton distribution functions (pdf), $\phi$, and the partonic cross section, $d{\hat \sigma}$:
\beq
d\sigma_{pp \to tqZ}=\sum_{a,b} \; 
\int dx_a \, dx_b \,  \phi_{a/p}(x_a, \mu_F) \, \phi_{b/p}(x_b, \mu_F) \, 
d{\hat \sigma}_{ab \to tqZ}(s_4, \mu_F) 
\eeq
with $\mu_F$ the factorization scale.

We then take Laplace transforms $d{\tilde{\hat\sigma}}_{ab \to tqZ}(N)=\int_0^s (ds_4/s) \,  e^{-N s_4/s} \, d{\hat\sigma}_{ab \to tqZ}(s_4)$, with $N$ the transform variable, and ${\tilde \phi}(N)=\int_0^1 e^{-N(1-x)} \phi(x) \, dx$.
Then, the cross section factorizes: 
\beq
d{\tilde \sigma}_{ab \to tqZ}(N)= {\tilde \phi}_{a/a}(N_a, \mu_F) \, {\tilde \phi}_{b/b}(N_b, \mu_F) \, d{\tilde{\hat \sigma}}_{ab \to tqZ}(N, \mu_F) \, .
\eeq

A further refactorization in terms of hard, $H$, and soft, $S$, functions as well as initial- and final-state functions $\psi$ and $J$, is written as
\beq
d{\tilde{\sigma}}_{ab \to tqZ}(N)={\tilde \psi}_{a/a}(N_a,\mu_F) \, {\tilde \psi}_{b/b}(N_b,\mu_F) \, {\tilde J}_q (N, \mu_F) \, {\rm tr} \left\{H_{ab \to tqZ} \left(\alpha_s(\mu_R)\right) \, {\tilde S}_{ab \to tqZ} \left(\frac{\sqrt{s}}{N \mu_F} \right)\right\} \, .
\eeq

Thus, comparing the previous two equations, we have
\beq
d{\tilde{\hat \sigma}}_{ab \to tqZ}(N,\mu_F)=
\frac{{\tilde \psi}_{a/a}(N_a, \mu_F) \, {\tilde \psi}_{b/b}(N_b, \mu_F) \, {\tilde J_q} (N, \mu_F)}{{\tilde \phi}_{a/a}(N_a, \mu_F) \, {\tilde \phi}_{b/b}(N_b, \mu_F)} \; \,  {\rm tr} \left\{H_{ab \to tqZ}\left(\alpha_s(\mu_R)\right) \, 
{\tilde S}_{ab \to tqZ}\left(\frac{\sqrt{s}}{N \mu_F} \right)\right\} \, .
\eeq

The resummed cross section is derived from the renormalization group evolution of the $N$-dependent functions above, and it is written as 
\beqa
d\tilde{\hat \sigma}_{ab \to tqZ}^{\rm resum}(N,\mu_F) & = &
\exp\left[\sum_{i=a,b} E_{i}(N_i)\right] \, 
\exp\left[\sum_{i=a,b} 2 \int_{\mu_F}^{\sqrt{s}} \frac{d\mu}{\mu} \gamma_{i/i}(N_i)\right] \, 
\exp\left[E'_q(N)\right]
\nonumber \\ && \hspace{-2mm}
\times {\rm tr} \left\{H_{ab \to tqZ}\left(\alpha_s(\sqrt{s})\right) {\bar P} \, \exp \left[\int_{\sqrt{s}}^{{\sqrt{s}}/N} \frac{d\mu}{\mu} \, \Gamma_{\! S \, ab \to tqZ}^{\dagger} \left(\alpha_s(\mu)\right)\right] \right.
\nonumber \\ && \hspace{3mm}
\left. \times {\tilde S}_{ab \to tqZ} \left(\alpha_s\left(\frac{\sqrt{s}}{N}\right)\right) \, P \, \exp \left[\int_{\sqrt{s}}^{{\sqrt{s}}/N} \frac{d\mu}{\mu} \, \Gamma_{\! S \, ab \to tqZ}
\left(\alpha_s(\mu)\right)\right] \! \right\} \, .
\eeqa
The first and third exponentials resum collinear and soft gluon emission from incoming partons and the outgoing light quark, respectively, while the second exponential gives the dependence on $\mu_F$ (see e.g. Ref. \cite{FK2020} for detailed expressions).
The soft anomalous dimensions $\Gamma_{\! S \, ab \to tqZ}$ for this process are $2 \times 2$ matrices and are known at one and two loops \cite{FK2020}, and partly at three loops \cite{NKloops}, and they are simply related to those for single-top processes \cite{NKsingletop,NK3loop}.

The expansion of the resummed cross section followed by inversion to momentum space provides a determination of soft-gluon NNLO corrections with no prescription needed. Thus, we derive aNNLO theoretical predictions for the differential cross section, with aNNLO = NLO + soft-gluon NNLO corrections.

\mysection{aNNLO cross sections for $tqZ$ and ${\bar t}qZ$ production}

In this section we present results for the total cross section for $tqZ$ production as well as for ${\bar t}qZ$ production through aNNLO. Our central results are given with a factorization and renormalization scale set to $m_t=172.5$ GeV. The NLO corrections are calculated using {\small \sc MadGraph5\_aMC@NLO} \cite{MG5}. We use MSHT20 pdf \cite{MSHT20} in our calculations.

We note that the soft-gluon contributions dominate the higher-order corrections, and at NLO they approximate very well the exact NLO cross section. This provides a strong reason for calculating them at higher orders.

\begin{table}[h]
\begin{center}
\begin{tabular}{|c|c|c|c|c|c|} \hline
\multicolumn{6}{|c|}{Sum of $tqZ$ and ${\bar t}qZ$ cross sections in $pp$ collisions at the LHC} \\ \hline
$\sigma$ in fb & 7 TeV & 8 TeV & 13 TeV & 13.6 TeV & 14 TeV \\ \hline
LO    & $153^{+2}_{-6}{}^{+3}_{-2}$ & $221^{+5}_{-11}{}^{+3}_{-4}$ & $741^{+34}_{-52}{}^{+9}_{-8}$ & $822^{+39}_{-59} \pm 9$ & $879^{+43}_{-65} \pm 9$ \\ \hline
NLO   & $165 \pm 3{}^{+2}_{-3}$ & $240^{+5}_{-3}{}^{+4}_{-3}$ & $850^{+19}_{-18}{}^{+11}_{-9}$ & $951^{+19}_{-21} \pm 11$ & $1022^{+24}_{-25}{}^{+12}_{-10}$ \\ \hline
aNNLO & $174^{+1}_{-3} \pm 3$ & $256^{+2}_{-3}{}^{+5}_{-3}$ & $908^{+6}_{-15}{}^{+10}_{-9} $ & $1012^{+6}_{-18}{}^{+10}_{-9}$ & $1087^{+7}_{-21}{}^{+12}_{-9}$ \\ \hline
\end{tabular}
\caption[]{The sum of the $tqZ$ and ${\bar t}qZ$ cross sections (in fb), with scale and pdf uncertainties, in $pp$ collisions with $\sqrt{S}=7$, 8, 13, 13.6, and 14 TeV, $m_t=172.5$ GeV, and MSHT20 pdf \cite{MSHT20}.}
\label{table}
\end{center}
\end{table}

In Table 1 we give results for the sum of the $tqZ$ and ${\bar t}qZ$ cross sections, together with scale and pdf uncertainties, for a variety of LHC energies at LO, NLO, and aNNLO using MSHT20 pdf at each order. We note that the scale uncertainty is reduced with each higher perturbative order, and at aNNLO it is comparable to the pdf uncertainty. We also note that much of the difference between the aNNLO and NLO results is due to differences in the pdf between orders. 

\begin{figure}[h]
\begin{center}
\includegraphics[width=88mm]{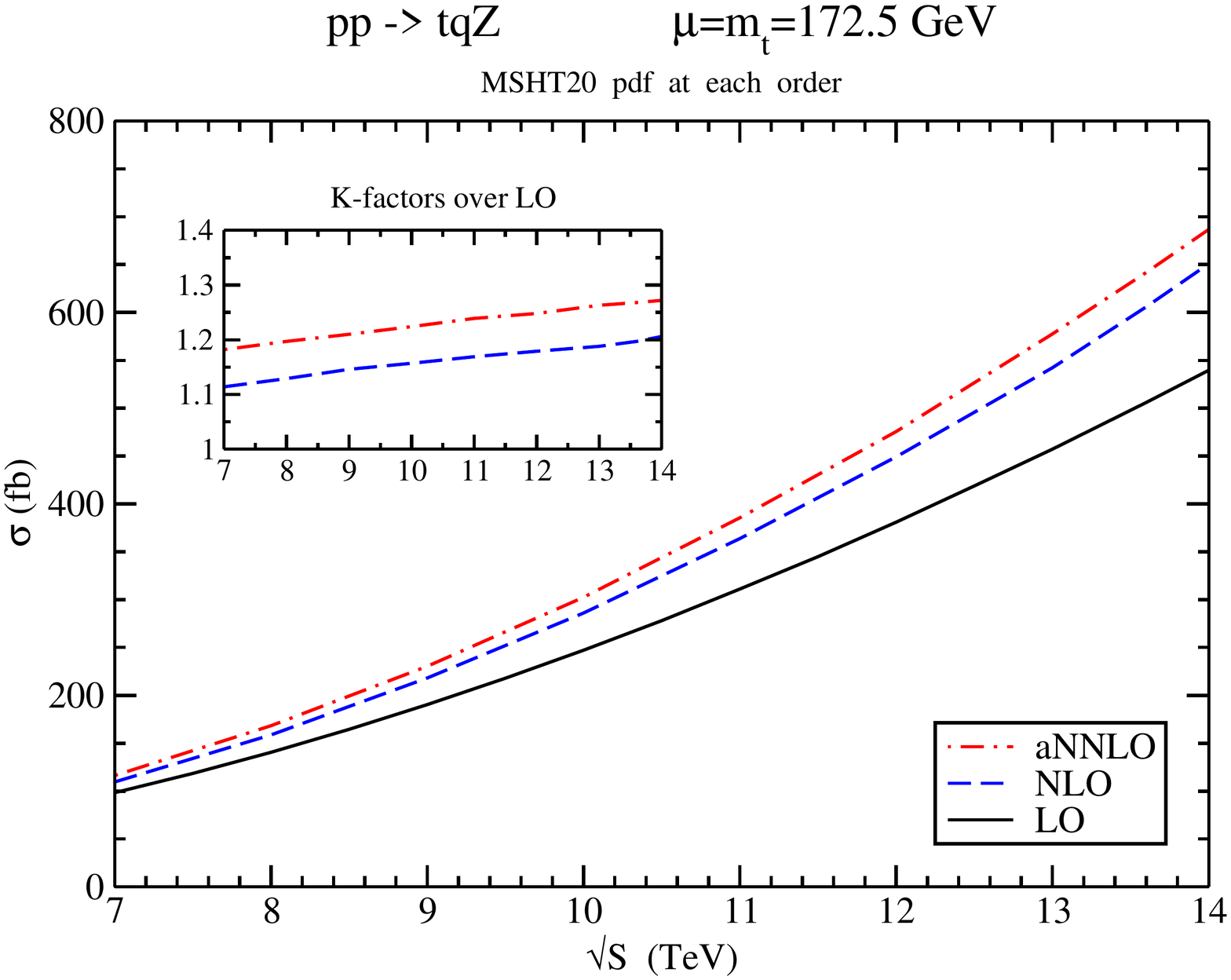} 
\includegraphics[width=88mm]{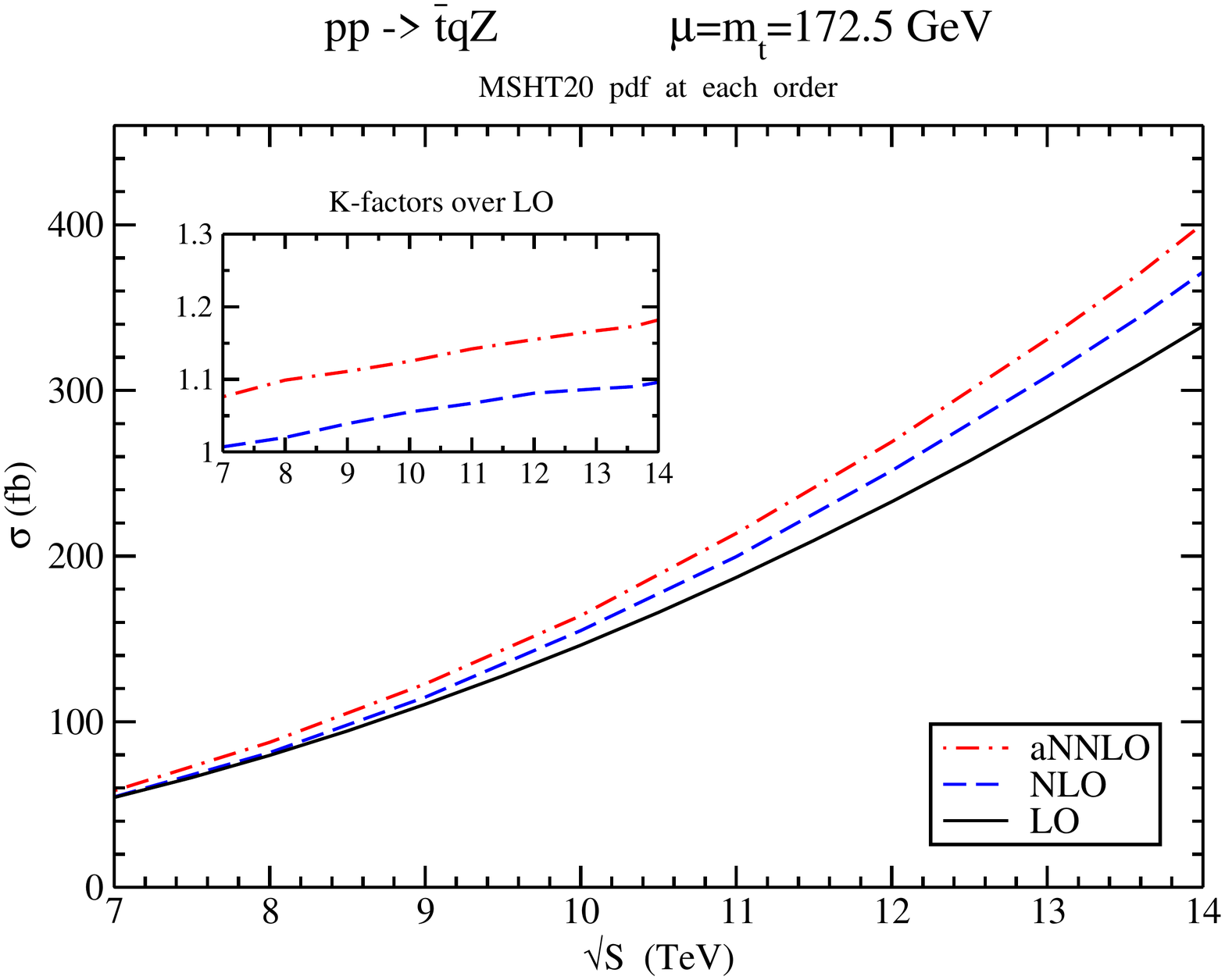} 
\caption{The $tqZ$ (left) and ${\bar t}qZ$ (right) production cross sections at LHC energies.}
\label{tqZ}
\end{center}
\end{figure}

In Fig. \ref{tqZ}, we plot the $tqZ$ and ${\bar t}qZ$ production cross sections at LHC energies. Results are shown at LO, NLO, and aNNLO. The $K$-factors, i.e. the ratios of the cross sections at different orders, are shown in the inset plots. Both the NLO/LO and the aNNLO/LO $K$-factors increase with collision energy, and they are not sensitive to possible cuts on the $p_T$ of the $Z$-boson. We find that the $K$-factors are larger for the $tqZ$ cross sections than those for ${\bar t}qZ$ production.

Overall, the aNNLO calculation provides an improved theoretical result with an increased cross section but with smaller scale uncertainty. At 13 TeV energy, the NLO corrections provide a 14.7\% increase over LO, and the aNNLO corrections provide an additional 7.8\% for a total of 22.5\% increase from the QCD corrections through aNNLO. At 13.6 TeV energy, the NLO corrections provide a 15.7\% increase over LO, and the aNNLO corrections provide an additional 7.4\% for a total of 23.1\% increase from the QCD corrections through aNNLO.

\mysection{Top-quark rapidity distributions in $tqZ$ production}

In this section, we present differential distributions in top-quark rapidity for $tqZ$ production.

\begin{figure}[h]
\begin{center}
\includegraphics[width=88mm]{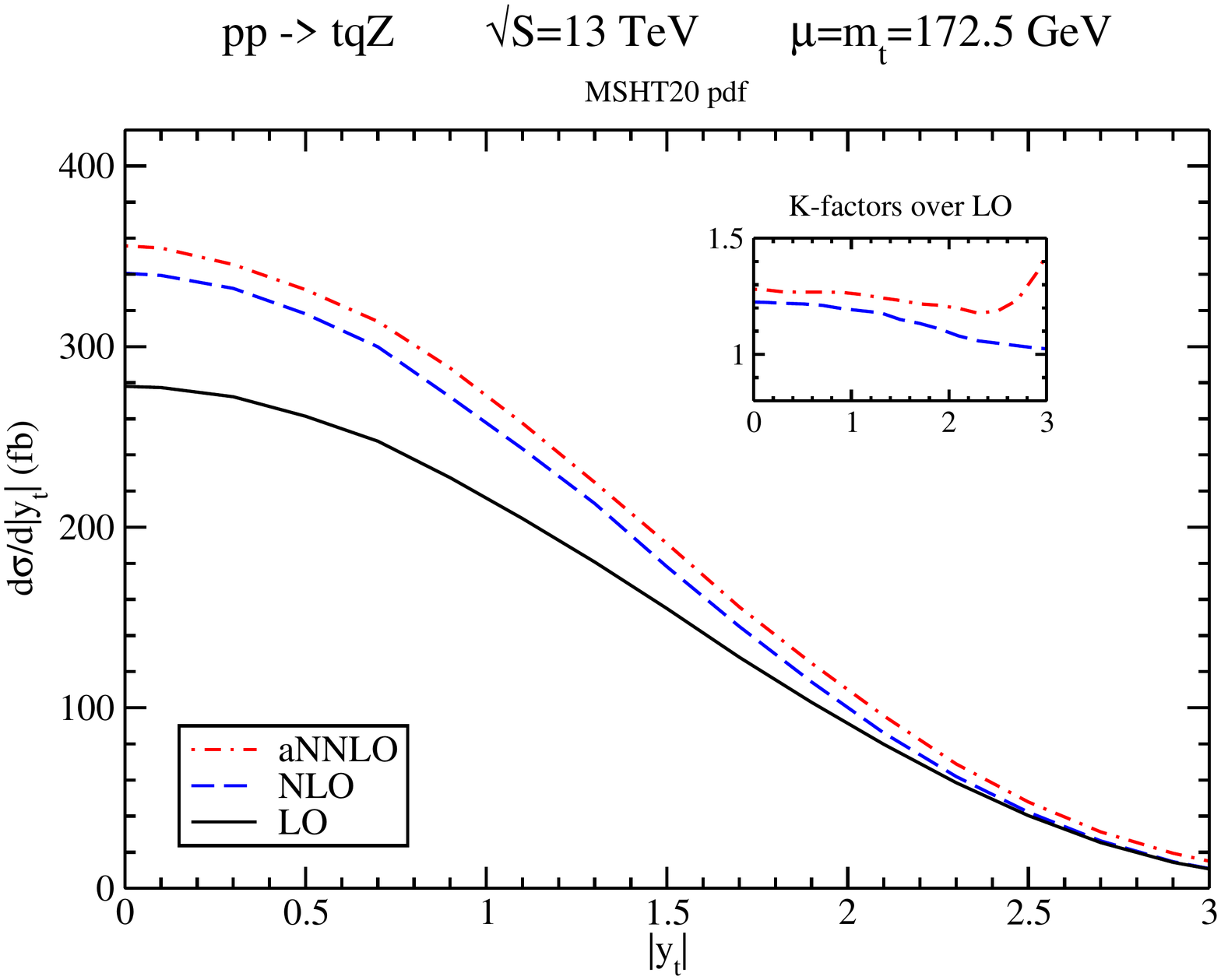}
\includegraphics[width=88mm]{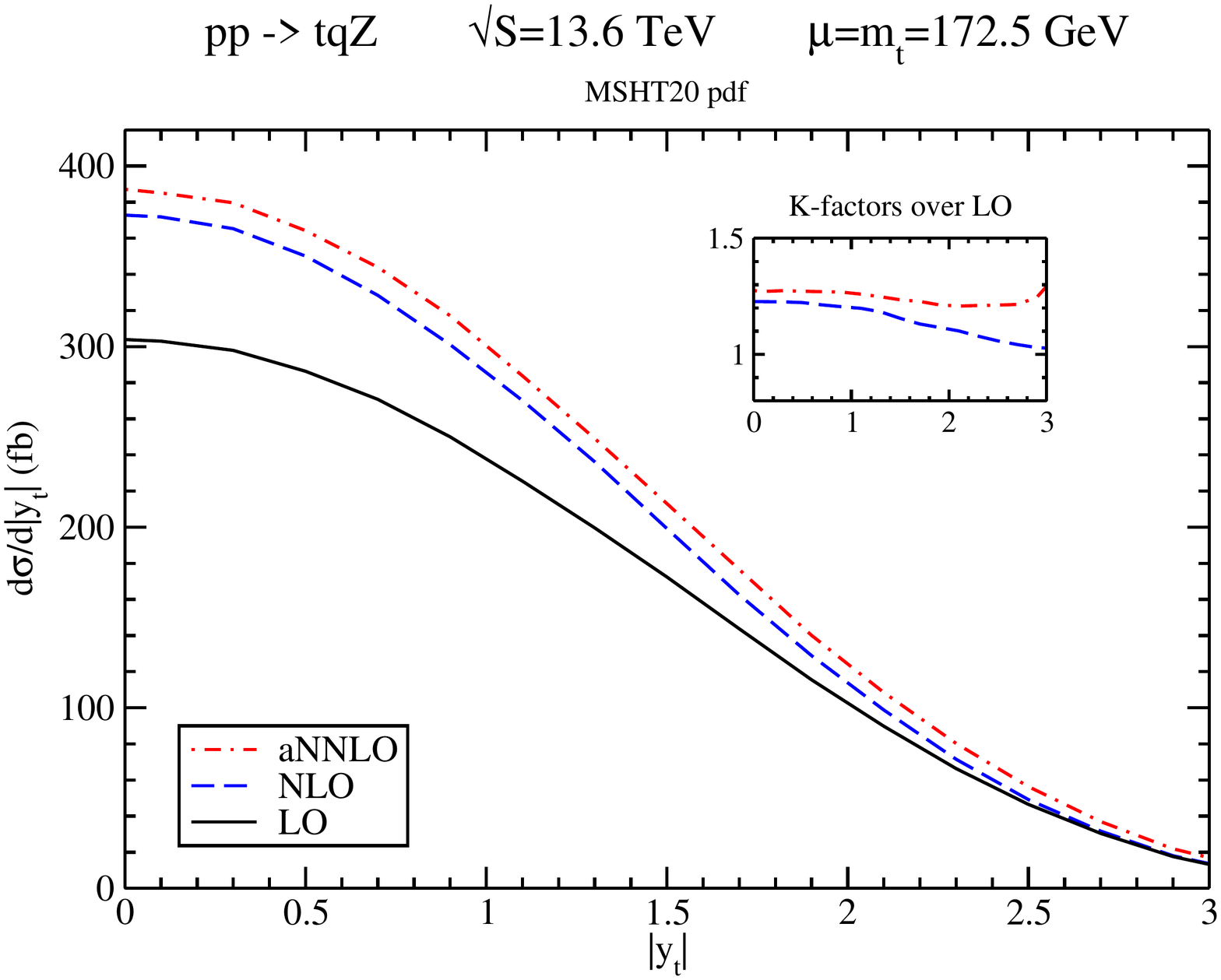}
\caption{The top-quark rapidity distributions in $tqZ$ production at 13 TeV (left) and 13.6 TeV (right) LHC energies.}
\label{yabstoptqZ}
\end{center}
\end{figure}

In Fig. \ref{yabstoptqZ}, we present our theoretical predictions for the top-quark rapidity distributions, specifically $d\sigma/d|y_t|$, at the LHC at 13 TeV and 13.6 TeV energies. Results through aNNLO are presented, and the inset plots display the $K$-factors over LO. 

We find significant enhancements from the aNNLO corrections, particularly at large rapidities. As for the total cross section, part of the difference between the aNNLO and the NLO distributions is a consequence of the order of the different pdf in the corresponding calculations.

We also note that the scale and pdf uncertainties get bigger at larger rapidities, $|y_t| > 2$. However, for small and moderate values of the top-quark rapidity, the scale and pdf uncertainties are very close to those for the total cross section. We also find similar conclusions for the top-quark rapidity distribution at 14 TeV LHC energy \cite{NKNYtqZ}.

\mysection{Summary}

We have studied $tqZ$ production in high-energy $pp$ collisions. We have presented detailed results for the aNNLO corrections from soft-gluon resummation for total cross sections and top-quark rapidity distributions at LHC energies. The NLO QCD corrections are significant, and the aNNLO contributions further improve the theoretical predictions.

\section*{Acknowledgements}
The work of N.K. is supported by the National Science Foundation under Grant No. PHY 2112025. N.Y. was supported by Daiko Foundation.


\begin{thebibliography}{99}

\bibitem{CMS1}
CMS Collaboration, JHEP {\bf 07}, 003 (2017) [arXiv:1702.01404].

\bibitem{ATLAS1}
ATLAS Collaboration, Phys. Lett. B {\bf 780}, 557 (2018) [arXiv:1710.03659].

\bibitem{CMS2}
CMS Collaboration, Phys. Lett. B {\bf 779}, 358 (2018) [arXiv:1712.02825].  

\bibitem{CMS3}
CMS Collaboration, Phys. Rev. Lett. {\bf 122}, 132003 (2019) [arXiv:1812.05900].

\bibitem{ATLAS2}
ATLAS Collaboration, JHEP {\bf 07}, 124 (2020) [arXiv:2002.07546].

\bibitem{CMS4}
CMS Collaboration, JHEP {\bf 02}, 107 (2022) [arXiv:2111.02860].

\bibitem{NKAB}
N. Kidonakis and A. Belyaev, JHEP {\bf 12}, 004 (2003) [arXiv:hep-ph/0310299].

\bibitem{LZLGZ}
B.H. Li, Y. Zhang, C.S. Li, J. Gao, and H.X. Zhu, Phys. Rev. D {\bf 83}, 114049 (2011) [arXiv:1103.5122].

\bibitem{NKtZ}
N. Kidonakis, Phys. Rev. D {\bf 97}, 034028 (2018) [arXiv:1712.01144].

\bibitem{MGNK}
M. Guzzi and N. Kidonakis, Eur. Phys. J. C {\bf 80}, 467 (2020) [arXiv:1904.10071].

\bibitem{YLSM}
Y.-B. Liu and S. Moretti, Chin. Phys. C {\bf 45}, 043110 (2021) [arXiv:2010.05148].

\bibitem{NKEM}
N. Kidonakis and E. Martin, Phys. Rev. D {\bf 90}, 054021 (2014) [arXiv:1404.7488]. 

\bibitem{CDVM}
V. Cirigliano, W. Dekens, J. de Vries, and E. Mereghetti, Phys. Rev. D {\bf 94}, 016002 (2016) [arXiv:1603.03049].

\bibitem{Yamanaka2017}
N. Yamanaka, B.K. Sahoo, N. Yoshinaga, T. Sato, K. Asahi, and B.P. Das, Eur. Phys. J. A {\bf 53}, 54 (2017) [arXiv:1703.01570].

\bibitem{CER}
J. Campbell, R.K. Ellis, and R. Rontsch, Phys. Rev. D {\bf 87}, 114006 (2013) [arXiv:1302.3856].

\bibitem{PTV}
D. Pagani, I. Tsinikos, and E. Vryonidou, JHEP {\bf 08}, 082 (2020) [arXiv:2006.10086].

\bibitem{DPS}
A. Denner, G. Pelliccioli, and C. Schwan, JHEP {\bf 10}, 125 (2022) [arXiv:2207.11264].

\bibitem{NKsingletop}
N. Kidonakis, Phys. Rev. D {\bf 74}, 114012 (2006) [arXiv:hep-ph/0609287]; D {\bf 75}, 071501 (2007) [arXiv:hep-ph/0701080]; D {\bf 81}, 054028 (2010) [arXiv:1001.5034]; D {\bf 82}, 054018 (2010) [arXiv:1005.4451]; D {\bf 83}, 091503 (2011) [arXiv:1103.2792]; D {\bf 88}, 031504 (2013) [arXiv:1306.3592]; D {\bf 93}, 054022 (2016) [arXiv:1510.06361]; D {\bf 96}, 034014 (2017) [arXiv:1612.06426].

\bibitem{NK3loop}
N. Kidonakis, Phys. Rev. D {\bf 99}, 074024 (2019) [arXiv:1901.09928].

\bibitem{NKNYtW}
N. Kidonakis and N. Yamanaka, JHEP {\bf 05}, 278 (2021) [arXiv:2102.11300].

\bibitem{NKttbar}
N. Kidonakis, Phys. Rev. D {\bf 82}, 114030 (2010) [arXiv:1009.4935]; D {\bf 84}, 011504 (2011) [arXiv:1105.5167]; D {\bf 90}, 014006 (2014) [arXiv:1405.7046];  D {\bf 91}, 031501 (2015) [arXiv:1411.2633]; D {\bf 91}, 071502 (2015) [arXiv:1501.01581]; D {\bf 101}, 074006 (2020) [arXiv:1912.10362]; in Snowmass 2021 [arXiv:2203.03698].

\bibitem{FK2020}
M. Forslund and N. Kidonakis, Phys. Rev. D {\bf 102}, 034006 (2020) [arXiv:2003.09021].

\bibitem{FK2021}
M. Forslund and N. Kidonakis, Phys. Rev. D {\bf 104}, 034024 (2021) [arXiv:2103.01228]; SciPost Phys. Proc. {\bf 8}, 021 (2022) [arXiv:2107.13489].

\bibitem{NKNYtqgamma}
N. Kidonakis and N. Yamanaka, Eur. Phys. J. C {\bf 82}, 670 (2022) [arXiv:2201.12877].

\bibitem{NKNYtqZ}
N. Kidonakis and N. Yamanaka, Phys. Lett. B {\bf 838}, 137708 (2023) [arXiv:2210.09542].

\bibitem{NKAT}
N. Kidonakis and A. Tonero, Phys. Rev. D {\bf 107}, 034013 (2023) [arXiv:2212.00096].

\bibitem{NKloops}
N. Kidonakis, SciPost Phys. Proc. {\bf 7}, 046 (2022) [arXiv:2109.14102]. 

\bibitem{MG5}
J. Alwall {\it et al.}, JHEP {\bf 07}, 079 (2014) [arXiv:1405.0301].

\bibitem{MSHT20}
S. Bailey, T. Cridge, L.A. Harland-Lang, A.D. Martin, and R.S. Thorne, Eur. Phys. J. C {\bf 81}, 341 (2021) [arXiv:2012.04684].


\end{thebibliography}
\end{document}